\documentclass[prd,superscriptaddress,showpacs,showkeys]{revtex4}
\usepackage{amsmath}
\usepackage{amssymb}
\usepackage{braket}
\usepackage{eurosym}
\usepackage{amsfonts}
\usepackage{graphicx}
\usepackage{calrsfs}
\usepackage[colorlinks=true,
            linkcolor=red,
            urlcolor=blue,
            citecolor=blue]{hyperref}

\def\be{\begin{equation}}
\def\ee{\end{equation}}
\usepackage{breakurl}

\begin{document}
\title{Stability of Effective Thin-shell Wormholes Under Lorentz Symmetry Breaking  Supported by  Dark Matter and Dark Energy}

\author{Ali \"{O}vg\"{u}n}
\email{ali.ovgun@pucv.cl}
\affiliation{Instituto de F\'{\i}sica, Pontificia Universidad Cat\'olica de
Valpara\'{\i}so, Casilla 4950, Valpara\'{\i}so, Chile}
\affiliation{Physics Department, Arts and Sciences Faculty, Eastern Mediterranean University, Famagusta, North Cyprus via Mersin 10, Turkey}
\affiliation{TH Division, Physics Department, CERN, CH-1211 Geneva 23, Switzerland }

\author{Kimet Jusufi}
\email{kimet.jusufi@unite.edu.mk}
\affiliation{Physics Department, State University of Tetovo, Ilinden Street nn,
1200, Macedonia}

\date{\today }

\begin{abstract}

In this paper, we construct generic, spherically symmetric  thin-shell wormholes and check their stabilities using the unified dark sector, including dark energy and dark matter. We give a master equation, from which one can recover, as a special case, other stability solutions for generic spherically symmetric thin-shell wormholes. In this context, we consider a particular solution; namely we construct an effective thin-shell wormhole under Lorentz symmetry breaking. We explore stability analyses using different models of the modified Chaplygin gas with constraints from cosmological observations such as seventh-year full Wilkinson microwave anisotropy probe data points, type Ia supernovae, and baryon acoustic oscillation. In all these models we find stable solutions by choosing suitable values for the parameters of the Lorentz symmetry breaking effect.  

\end{abstract}


\maketitle
 
\section{Introduction}

Since the important work of Morris and Thorne on traversable wormholes \cite{moris,moris2}, explorations of stable wormhole solutions have become a hot topic of research  \cite{visser1,visser2,visser3,visser4, Lobo,eiroa1,Musgrave:1995ka,Musgrave:1997sfw,lake,eiroa2,ao1,ao2,rahaman,ao3,kimett,ao4,ao5,prda,ayan2,sharif1,sharif2,sharif3,sharif4,varela,kuh,azam,rah,sharif5,ayan,ao6,ao7,wang1,wang2,de,k,eid1,ne,ak,eid2,lobo2,ef,can,pitell,wangm,wang11,wang22,Richarte:2017iit,Zaeem-ul-HaqBhatti:2017kaw,Bhatti:2017yeg,Guendelman:2016bwj,Guendelman:2015wsv}. However, there are conceptual problems related to these exotic objects; for example, the main problem with stable wormholes is that a kind of exotic matter is needed at the throat of a wormhole that connects two different regions of space-time \cite{visser2}. According to the general theory of relativity, wormholes with exotic matter do not satisfy the null energy condition \cite{visser1}.

Within the context of the general theory of relativity, the first matching of thin-shells was studied in 1924 by Sen \cite{sen}, then  by Lanczos in 1924 \cite{lanczos}, Darmois in 1927 \cite{darmois}, and Israel in 1966 \cite{israel}. Israel produced a cut-and-paste technique by applying the Gauss-Codacci equations to a non-null 3D hypersurface imbedded in a 4D space time \cite{israel,Guendelman:2008ip}. Then, Visser used this thin-shell formalism to construct a thin-shell wormhole (TSW), but to open the throat of the wormhole and make it stable, exotic matter is required \cite{visser1,visser2,visser3,visser4}. For this purpose, many kinds of energy of states (EOS) are used that are also familiar to us from cosmological models, such as Chaplygin gas \cite{ref:CG}, generalized Chaplygin gas \cite{ref:GCG}, modified Chaplygin gas (MCG) \cite{ref:MCG}, barotropic fluids \cite{varela}, logarithmic gas, etc. \cite{ref:LuMCG,ao1,ao2,ao3,ao4,ao5,ao6,ao7}. The increasing number and precision of cosmological experiments has revealed  more evidence of the standard $\Lambda$CDM model, which is an accurate description of gravity on cosmological scales \cite{suzuki,anderson,parkinson,hinshaw,ade,adee}. The main ingredients of this model are the cosmological constant $\Lambda$, which is needed to explain the acceleration of the Universe and is also known as Dark Energy (DE) \cite{Ovgun:2017iwg,Ovgun:2016oit}, and the nonrelativistic dark matter (DM) component, which explains the observed rotational velocity curves of galaxies \cite{cosmo1,cosmo2}. The other way to explain DE is to define a single perfect fluid with a constant negative pressure. The simplest unified dark energy model can be used to understand $\Lambda$CDM such that the DM and DE are used as a single unified DE fluid such as a MCG model which, is an unified dark sector \cite{de2,avelino,bento,bilic,gorini}. The MCG model has two components: DM is presented by $a^{-3}$ and the remaining part is DE. The best-fit values for this model according to the seventh-year Wilkinson microwave anisotropy probe (WMAP) and the sloan digital sky survey (SDSS)'s baryon oscillation spectroscopic survey (BAO) are as follows: $B_s=0.822$, $\alpha=1.724$ and $B=-0.085$ \cite{de,ref:Lu}. There are further constraints from the cosmic microwave background (CMB) shift parameters (BAO, type Ia supernova data (SN Ia)), observational Hubble data, and cluster X-ray gas mass fraction (CBF) as follows: $B_{s}=0.7788^{+0.0736}_{-0.0723}$ ($1\sigma$) $^{+0.0918}_{-0.0904}$ $(2\sigma)$, $\alpha=0.1079^{+0.3397}_{-0.2539}$ ($1\sigma$) $^{+0.4678}_{-0.2911}$ $(2\sigma)$, $B=0.00189^{+0.00583}_{-0.00756}$ ($1\sigma$) $^{+0.00660}_{-0.00915}$ $(2\sigma)$. Wang and Meng studied wormholes in the context of modified gravities through observations \cite{wang11}, and in their second work, they explored model-independent wormholes for the first time based on Gaussian Processes \cite{wang22}. In this paper, our aim is to construct an effective TSW with DE and DM in the background evolution by using BAO, SN Ia, and CMB data \cite{ref:LuMCG}. 

In this paper, we use a Schwarzschild black hole solution under the Lorentz symmetry breaking recently investigated by Betschart et al. \cite{kant} to construct an effective TSW. Betschart et al. \cite{kant} starting from a nonbirefringent modified Maxwell theory, derived three effective metrics in a Schwarzschild background which shall be used in this work. A  similar study has been done in the context of cosmic string space-time for the scalar spin zero particles subject to a scalar potential \cite{bak1,bak2,bak3,bak4,bak5,bakke0,bakke1,hof}.  Furthermore it is used two possible models of the anisotropy that is generated under the Lorentz symmetry breaking effect. They showed the change in the cosmic string space-time under the effects of the Lorentz symmetry violation using the modified mass term in the scalar potential.

This paper is organized as follows: In Section II, we review the generic, spherical symmetric TSW. In Section III, we define the stability conditions for the TSW. In Section IV, we give an example of an effective TSW using the formalism that we defined in a previous section and check its stabilities via different types of fluids. Then, we use the constrained cosmological parameter to study the effective TSW. In Section V, we discuss our results.   

\section{General Analysis of TSW}

\begin{figure}[!htb]
  \centering
\includegraphics[width=3 in]{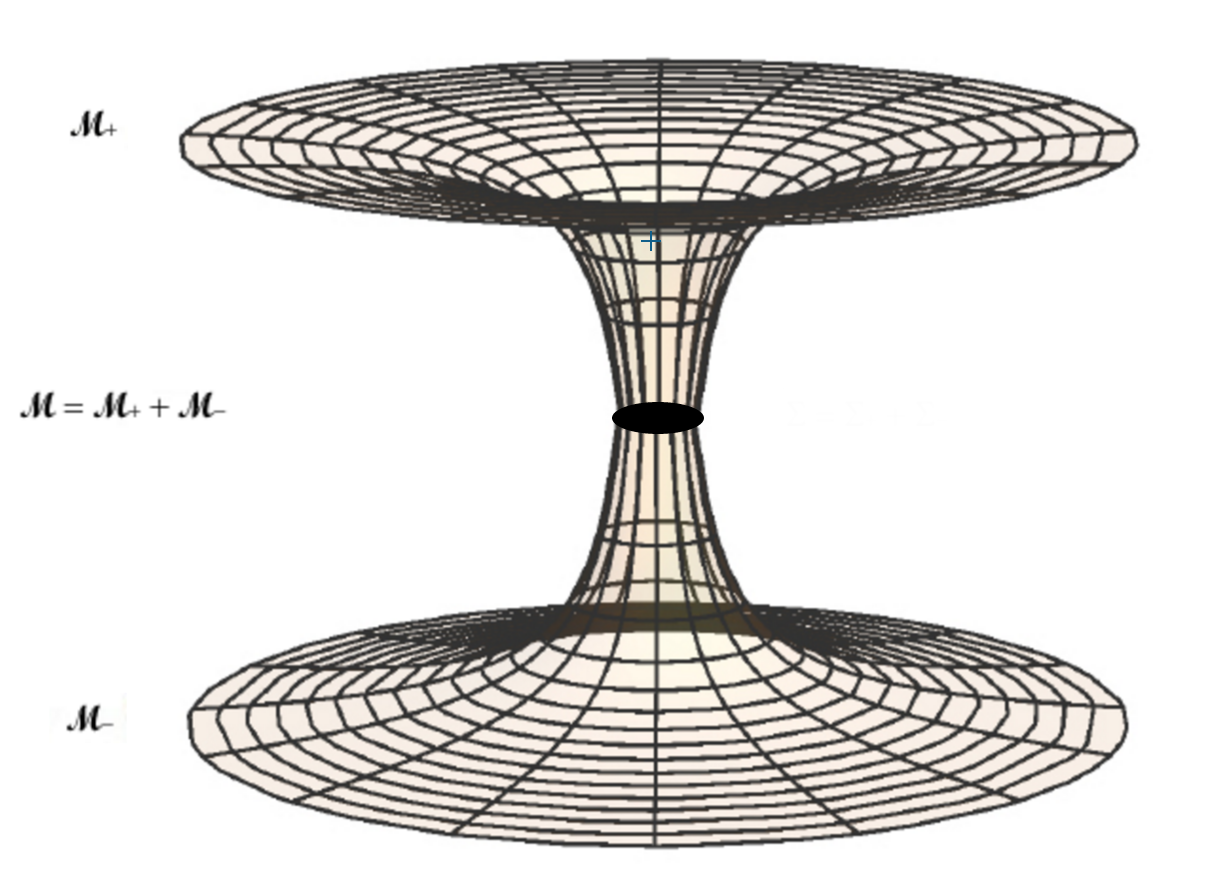}
  \caption{The figure shows a diagram of a TSW.}
  \label{figwh}
\end{figure}

To construct a stable TSW, we use two copies of the spherically symmetric geometries as follows:
\begin{equation}
ds_{\pm}^2=-F_{\pm}(r)dt^2+\frac{dr^2}{G_{\pm}(r)}+r^2 d\theta^2+r^2 \sin^2 \theta d\phi^2.\label{effmetric1}
\end{equation}
It is noted that two distinct geometries have manifolds of ${ M_+}$ and ${ M_-}$. The metrics of the manifolds are given by $g_{\mu \nu}^+(x^{\mu}_+)$ and $g_{\mu \nu}^-(x^{\mu}_-)$, which the coordinate systems $x^{\mu}_+$ and $x^{\mu}_-$ are defined independently. Our aim is to obtain a single manifold
${M}$ from the manifolds ${ M_+}$ and 
${M_-}$ using the copy and paste technique. For this purpose, the boundaries are defined as follows:  $\Sigma=\Sigma_+=\Sigma_-$ \cite{Lobo}. This construction is depicted in Fig. \ref{figwh}. 

Then let us proceed with the cut and paste technique to construct a TSW using the metric \eqref{effmetric1} and choosing two identical regions 
\begin{equation}
 M^{(\pm)}=\left\lbrace r^{(\pm)}\leq a,\,\,a>r_{H}\right\rbrace ,
\end{equation}
where the radius of the throat $a$ should be greater than the radius of the event horizon $r_{h}$. After we glue these regular regions at the boundary hypersurface $\Sigma^{(\pm)}=\left\lbrace r^{(\pm)}=a,a>r_{H}\right\rbrace $, we end up with a complete manifold $M=M^{+}\bigcup M^{-}$.

In accordance with the Darmois\textendash Israel formalism, the coordinates on $M$ are chosen as $x^{\alpha}=(t,r,\theta,\phi)$. On the other hand, we write the induced metric $\Sigma$ with the coordinates $\xi^{i}=(\tau,\theta,\phi)$, which are related to the coordinates of $M$, after using the following coordinate transformation
\begin{equation}
g_{ij}=\frac{\partial x^\alpha}{\partial \xi^i}\frac{\partial x^\beta}{\partial \xi^j}g_{\alpha \beta}. \label{10}
\end{equation}

Then we write the parametric equation for the boundary hypersurface on the induced metric $\Sigma$ as follows:
\begin{equation}
\Sigma: H(r,\tau)=r-a(\tau)=0.
\end{equation}

Note that in order to study the dynamics of the induced metric $\Sigma$, we let the radius of the throat of TSW to be time dependent by incorporating the proper time on the shell i.e., $a=a(\tau)$. Then the induced metric is obtained as follows:
\begin{equation}
\mathrm{d}s_{\Sigma}^{2}=-\mathrm{d}\tau^{2}+a(\tau)^{2}\left(\mathrm{d}\theta^{2}+\sin^{2}\theta\,\mathrm{d}\phi^{2}\right).\label{metric}
\end{equation}

The Darmois-Israel junction conditions from the Lanczos equation \cite{Musgrave:1995ka,Musgrave:1997sfw} are used to glue two manifolds at the boundary hypersurface $\Sigma$ where the field equations projected on the shell as follows:
\begin{equation}
{S^{i}}_{j}=-\frac{1}{8\pi}\left(\left[{K^{i}}_{j}\right]-{\delta^{i}}_{j}\,K\right),  \label{darmois}
\end{equation}
where ${S^{i}}_{j}=diag(-\sigma,p_{\theta},p_{\phi})$
is the surface stress-energy tensor of $\Sigma$ and the trace of the extrinsic curvature is calculated as $K=trace\,[{K^{i}}_{i}]$. Moreover, the extrinsic curvature $K$ is not continuous across $\Sigma$ so we define the discontinuity as $[K_{ij}]={K_{ij}}^{+}-{K_{ij}}^{-}$.  The expression of the extrinsic curvature $K_{ij}$ is written as follows:
\begin{equation}
K_{ij}^{(\pm)}=-n_{\mu}^{(\pm)}\left(\frac{\partial^{2}x^{\mu}}{\partial\xi^{i}\partial\xi^{j}}+\Gamma_{\alpha\beta}^{\mu}\frac{\partial x^{\alpha}}{\partial\xi^{i}}\frac{\partial x^{\beta}}{\partial\xi^{j}}\right)_{\Sigma}.
\end{equation}

The unit vectors ${n_{\mu}}^{(\pm)}$, which are normal to $M^{(\pm)}$, are chosen as
\begin{equation}
n_{\mu}^{(\pm)}=\pm\left(\left\vert g^{\alpha\beta}\frac{\partial H}{\partial x^{\alpha}}\frac{\partial H}{\partial x^{\beta}}\right\vert ^{-1/2}\frac{\partial H}{\partial x^{\mu}}\right)_{\Sigma}.
\end{equation}

Note that the prime and the dot represent the derivatives with respect
to $r$ and $\tau$, respectively. Then we obtain the following components of the extrinsic curvature: 

\begin{eqnarray}
{{K^{\tau}}_{\tau}}^{(\pm)}=\pm \frac{\dot{a}^2 F G'-\dot{a}^2G F'-2 F G \ddot{a}-G^2 F'}{2 F \sqrt{F G(\dot{a}^2 +G)}},
\end{eqnarray}
\begin{equation}
{{K^\theta}_\theta}^{(\pm)}={{K^\phi}_\phi}^{(\pm)}=\mp {\frac { \sqrt{F G \left( \dot{a}^2 +G\right)}}{a  F  }}.
\end{equation}
Note that for a given radius $a$, the energy density on the shell is
$\sigma$, while the pressure $p$ is $p=p_{{\theta}}=p_{{\phi}}$.
After algebraic manipulation, we calculate the energy density $\sigma$
\begin{eqnarray}
\sigma=-\frac{1}{2}{\frac {\sqrt {F  G  \left( \dot{a}^2+G  \right) }}{\pi F a }}
,\label{255}
\end{eqnarray}

and the surface pressure $p$
\begin{eqnarray}
p=\frac{1}{8} \frac{a ( a G
 F'\dot{a}^2-a  F
 G' \dot{a}^2+2\,a G F  \ddot{a}   +a F' G^{2}+2\,G F
   \dot{a}^2+2\,F   G^{2} )}{\sqrt {F G \dot{a}^2+G }F \pi}
.\label{266}
\end{eqnarray}

We consider the value of the radius of throat as a constant value $a_{0}$ to study the mechanical stability of the TSW so that the first and second derivatives of the $a_{0}$ become zero $\dot{a}=0$ and $\ddot{a}=0$. The resulting energy density in static configuration is 
\begin{eqnarray}
\sigma_{0}=-\frac{1}{2 \pi a}{\frac {G \sqrt{F }}{F }}\label{277},
\end{eqnarray}
and similarly the surface pressure is
\begin{eqnarray}
p_{0}=\frac{1}{8 \pi} \frac{a (a F'  G^{2}+2\,F   G^{2} )}{\sqrt {G}F }.\label{288}
\end{eqnarray}

\section{Stability of the TSW}
In this section we study the stability of the TSW. Using the following conservation law for the wormhole's throat
\be
-\nabla_i S^i_j = \left[T_{\alpha \beta} \frac{\partial x^\alpha}{\partial \xi^j}
  n^\beta \right],
\label{ceq}
\ee
the equation of the energy conservation is obtained as follows: 
\begin{eqnarray}
\frac{d}{d\tau}\left(\sigma A\right)+p\frac{d A}{d\tau}=0,\label{34}
\end{eqnarray}
where $A$ is the area of the wormhole's throat. After we rearrange Eq.(\ref{255}), the equation for the classical motion is obtained as follows: 
\begin{eqnarray}
\dot{a}^{2}=-V(a),\label{35}
\end{eqnarray}
where the potential $V(a)$ is 
\begin{eqnarray}
V(a)=\frac{4 F \pi^2 a^2 \sigma^2 - G^2}{G}.\label{36}
\end{eqnarray}

In order to investigate the stability of TSW, we expand the potential
$V(a)$ around the static solution, where we assume that the throat of wormhole is at rest,
\begin{eqnarray}
V(a)=V(a_{0})+V^{\prime}(a_{0})(a-a_{0})+\frac{V^{\prime\prime}(a_{0})}{2}(a-a_{0})^{2}+O(a-a_{0})^{3}.\label{37}
\end{eqnarray}

Then we calculate the second derivative of the potential $V''(a_{0})$ as follows: 

\begin{eqnarray}
V''(a_{0}) & = & \frac{16\pi}{a^{2}FG^{2}}\,\left(\left\{ aFG'-G\left[aF'-\left(\psi'+\frac{3}{2}\right)F\right]\right\} \psi a\sqrt{FG^{2}}-\frac{1}{16}\,G\Xi\right),\label{v2s}
\end{eqnarray}
with
\begin{equation}
\Xi=\left(a^{2}F''G^{2}-2\,a^{2}G''FG+2\,a^{2}FG'-2\,aG\left(aG'-2\,F\right)G'-4\,aG^{2}F'+32\,a^{2}\pi^{2}F^{2}\psi^{2}+8\,G^{2}\left(\psi'+\frac{3}{4}\right)F\right),
\end{equation}
where we introduce $\psi^{\prime}=p'/\sigma'$. The TSW is stable if and only if the second derivative of the potential is positive $V^{\prime\prime}(a_{0})\geq0$ under radial perturbations. The equation of motion of the throat, for a small perturbation becomes
\begin{eqnarray}
\dot{a}^{2}+\frac{V^{\prime\prime}(a_{0})}{2}(a-a_{0})^{2}=0.
\end{eqnarray}
Noted that for the condition of $V^{\prime\prime}(a_{0})\geq0$, TSW
is stable where the motion of the throat is oscillatory with angular
frequency $\omega=\sqrt{\frac{V^{\prime\prime}(a_{0})}{2}}$. In the next section, we construct an effective TSW using different types of fluids and we check its mechanical stability.

\section{Example: An Effective TSW}
In this section, we construct an effective TSW using the effective Schwarzschild spacetime metric under the Lorentz symmetry breaking which was studied by Betschart et al. \cite{kant}. They first studied the Ricci-flat effective metric with changed horizon as follows:
\begin{equation}
ds^2=-\left(1-\frac{2M}{r}\right)dt^2+\left(1-\frac{2M}{r}\right)^{-1}dr^2+r^2 d\theta^2+r^2 \sin^2 \theta d\varphi^2,\label{met0}
\end{equation}
where the mass is $M=m(1+\epsilon_1)$. Then the second effective metric, or the non-Ricci-flat effective metric with unchanged horizon was found as follows:
\begin{equation}
ds^2=-\left(1-\frac{2m}{r}\right)dt^2+\frac{1}{1-\eta}\left(1-\frac{2m}{r}\right)^{-1}dr^2+r^2 d\theta^2+r^2 \sin^2 \theta d\varphi^2.\label{met1}
\end{equation}

Last, the Ricci-flat effective metric with an unchanged horizon is obtained by \cite{kant} as follows:
\begin{equation}
ds^2=-\frac{1}{1+\epsilon_2}\left(1-\frac{2m}{r}\right)dt^2+\left(1-\frac{2m}{r}\right)^{-1}dr^2+r^2 d\theta^2+r^2 \sin^2 \theta d\varphi^2.\label{met2}
\end{equation}

We use this black hole's solution to construct an effective TSW, but first we rewrite these three separate cases in a single and a more compact form
\begin{equation}
ds^2=-F(r)dt^2+\frac{1}{G(r)}dr^2+r^2 d\theta^2+r^2 \sin^2 \theta d\varphi^2,\label{effmetric}
\end{equation}
where
\begin{equation}
F(r)=\frac{1}{1+\alpha_1}\left(1-\frac{2 M}{r}\right),\label{effmetric11}
\end{equation}
\begin{equation}
G(r)=\left(1-\alpha_2 \right)\left(1-\frac{2 M}{r}\right).\label{effmetric2}
\end{equation}

It is not difficult to see that Eq. (\ref{effmetric}) reduces to Eq. (\ref{met2}) for the $\alpha_1=\epsilon_2$, and $\epsilon_1=\alpha_2=0$. Setting $\epsilon_1=\alpha_1=0$, and $\alpha_2=\eta$, we find the metric (\ref{met1}). Moreover, for $\alpha_1=\alpha_2=0$, and $M=m(1+\epsilon_1)$ we recover the metric (\ref{met0}). After we use the Darmois-Israel junction conditions \ref{darmois} for the Schwarzschild spacetime under the Lorentz symmetry breaking \eqref{effmetric}, we obtain the resulting energy density 
\begin{eqnarray}
\sigma={\frac{1+\alpha_{1}}{\left(-2\,a+4\,M\right)\pi}\sqrt{-8\,{\frac{\left(-\frac{a}{2}+M\right)^{2}\left(\left(\frac{1}{2}\,{\it \dot{a}}^{2}-\frac{\alpha_{2}}{2}+\frac{1}{2}\right)a+M\left(\alpha_{2}-1\right)\right)\left(\alpha_{2}-1\right)}{\left(1+\alpha_{1}\right){a}^{3}}}}},\label{25}
\end{eqnarray}
and the surface pressure 
\begin{eqnarray}
p={\frac{1}{a\left(-a+2\,M\right)\pi}\frac{\left(\Upsilon\right)\left(\alpha_{2}-1\right)}{\sqrt{-{\frac{\left(-a+2\,M\right)^{2}\left(\alpha_{2}-1\right)\left({\it \dot{a}}^{2}a+2\,\alpha_{2}M-\alpha_{2}a-2\,M+a\right)}{\left(1+\alpha_{1}\right){a}^{3}}}}}},\label{26}
\end{eqnarray}

\[
\Upsilon=\frac{1}{4}\,a^{4}{\it \ddot{a}}+\left(-M{\it {\it \ddot{a}}}+\frac{1}{4}\,{\it \dot{a}}^{2}-\frac{\alpha_{2}}{4}+\frac{1}{4}\right)a^{3}+\frac{5}{4}\,\left(\frac{4}{5}M{\it {\it \ddot{a}}}-\frac{4}{5}\,{\it {\it \dot{a}}}^{2}+\alpha_{2}+\frac{{\it {\it \ddot{a}}}}{5}-1\right)Ma^{2}-2\,\left(-\frac{1}{2}\,{\it {\it \dot{a}}}^{2}+\alpha_{2}+\frac{{\it \ddot{a}}}{4}-1\right)M^{2}a
\]

\[\times \left(M\left(\alpha_{2}-1\right)-\frac{1}{2}\,{\it {\it \ddot{a}}}^{2}\right)M^{2}.\]

At this point, let us briefly mention that since we study
the wormhole stability at a static configuration, we use the radius of the throat $a$ as a constant $a_{0}$ so that there is a vanishing acceleration $\dot{a}=0$,
and $\ddot{a}=0$. The energy density and the surface pressure for the effective TSW are obtained as follows:
\begin{eqnarray}
\sigma_{0}=-\frac{1}{2}\,{\frac{1+\alpha_{1}}{a\pi}\sqrt{{\frac{\left(1-\alpha_{2}\right)^{2}}{1+\alpha_{1}}\left(1-2\,\frac{M}{a}\right)}}},\label{27}
\end{eqnarray}

\begin{eqnarray}
p_{0}=\frac{1}{8}{\frac{a\left(1+\alpha_{1}\right)}{\pi}\left(2\,{\frac{M\left(1-\alpha_{2}\right)^{2}}{a\left(1+\alpha_{1}\right)}\left(1-2\,\frac{M}{a}\right)}+2\,{\frac{\left(1-\alpha_{2}\right)^{2}}{1+\alpha_{1}}\left(1-2\,\frac{M}{a}\right)^{2}}\right){\frac{1}{\sqrt{{\frac{\left(1-\alpha_{2}\right)^{2}}{1+\alpha_{1}}\left(1-2\,{\frac{M}{a}}\right)^{3}}}}}}.\label{28}
\end{eqnarray}

It's obvious from Eq. \eqref{27} that the energy density is negative, i.e. $\sigma_{0}<0$,
which implies that the weak and dominant energy conditions are violated.

\subsection{Stability analysis of the effective TSW with MCG}

We consider the DE as an exotic matter on the shell to open the throat of the wormhole. According to the MCG \cite{ref:MCG}, we can model it with the following EOS:

\begin{equation}
\psi=\omega\left(\frac{1}{\sigma}-\frac{1}{\sigma_{0}}\right)+p_{0}, \label{MCG1}
\end{equation}
which is one of the possible candidate for the acceleration of the universe. Then we calculate the derivative of Eq. (\ref{MCG1}) as follows:
\begin{equation}
\psi^{\prime}(\sigma_{0})=-\frac{\omega}{\sigma_{0}^{2}}.
\end{equation}

The second derivative of the potential respect to $a$ is obtained as follows:
\begin{eqnarray}
V''(a) & = & {\frac{\varTheta(a)}{\left(-a+2\,M\right)\left(-1+\alpha_{2}\right)a^{3}\left(1+\alpha_{1}\right)}}, \label{V2MCG}
\end{eqnarray}
where
\[
\varTheta=\left(\left(-2\,\alpha_{1}-2\right)\alpha_{2}^{2}+\left(4\,\alpha_{1}+4\right)\alpha_{2}-16\,\pi^{2}\omega-2\,\alpha_{1}-2\right)a^{6}+4\,\left(\left(1+\alpha_{1}\right)\alpha_{2}^{2}+\left(-2\,\alpha_{1}-2\right)\alpha_{2}+4\,\pi^{2}\omega+\alpha_{1}+1\right)Ma^{5}
\]

\[
+\left(-2\,\left(-1+\alpha_{2}\right)^{2}M^{2}\left(1+\alpha_{1}\right)+\left(6\,\alpha_{1}+6\right)\alpha_{2}^{2}+\left(-12\,\alpha_{1}-12\right)\alpha_{2}+32\,\pi^{2}\omega+6\,\alpha_{1}+6\right)a^{4}
\]

\[
-18\,\left(\left(1+\alpha_{1}\right)\alpha_{2}^{2}+\left(-2\,\alpha_{1}-2\right)\alpha_{2}+{\frac{32\,\pi^{2}\omega}{9}}+\alpha_{1}+1\right)Ma^{3}+12\,\left(1+\alpha_{1}\right)\left(-1+\alpha_{2}\right)^{2}\left(M^{2}-\frac{1}{2}\right)a^{2}
\]

\[+20\,M\left(-1+\alpha_{2}\right)^{2}\left(1+\alpha_{1}\right)a-16\,\left(-1+\alpha_{2}\right)^{2}M^{2}\left(1+\alpha_{1}\right),\]

and then the $\omega$ for the case of $V''(a_{0})\geq 0$ is obtained as follows:
\begin{eqnarray}
\omega=\frac{1}{8}\,{\frac{\left(\alpha_{2}-1\right)^{2}\left(a^{6}-2\,Ma^{5}+\left(M^{2}-3\right)a^{4}+9\,Ma^{3}+\left(-6\,M^{2}+3\right)a^{2}-10\,Ma+8\,M^{2}\right)\left(1+\alpha_{1}\right)}{\pi^{2}a^{3}\left(Ma^{2}-a^{3}-4\,M+2\,a\right)}}. \label{omega1}
\end{eqnarray}
 
The stability regions of the effective TSW with MCG is shown on plots of $\omega$ versus $a_{0}$ for
different values of the parameters $\epsilon_{1}$, $\alpha_{1}$ and $\alpha_{2}$ in Figs. (2, 4, 6).  Then, we show the effect of $\alpha_{1}$, $\alpha_{2}$, $\epsilon_{1}$ on the second derivative of the potential, where it is stable at $V''(a_{0})\geq 0$ and outside the event horizon $a>r_{H}$, by plotting in Figs. (3, 5, 7).

\begin{figure}[!htb]

  \includegraphics[width=4 in]{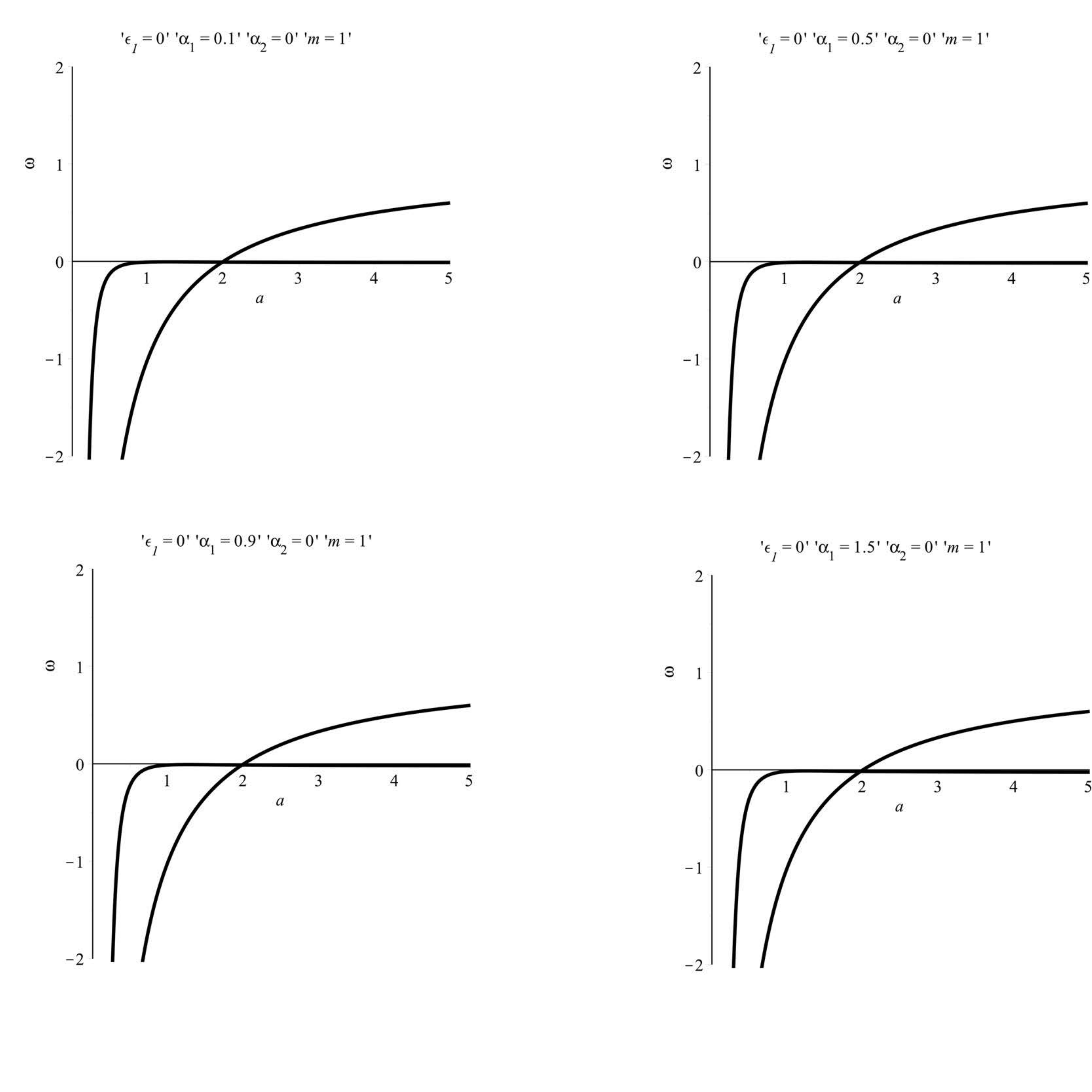} %
\caption{Here we plot the $\omega$ versus $a$ to show  stability regions via the MCG}
\end{figure}

\begin{figure}[!htb]

  \includegraphics[width=4 in]{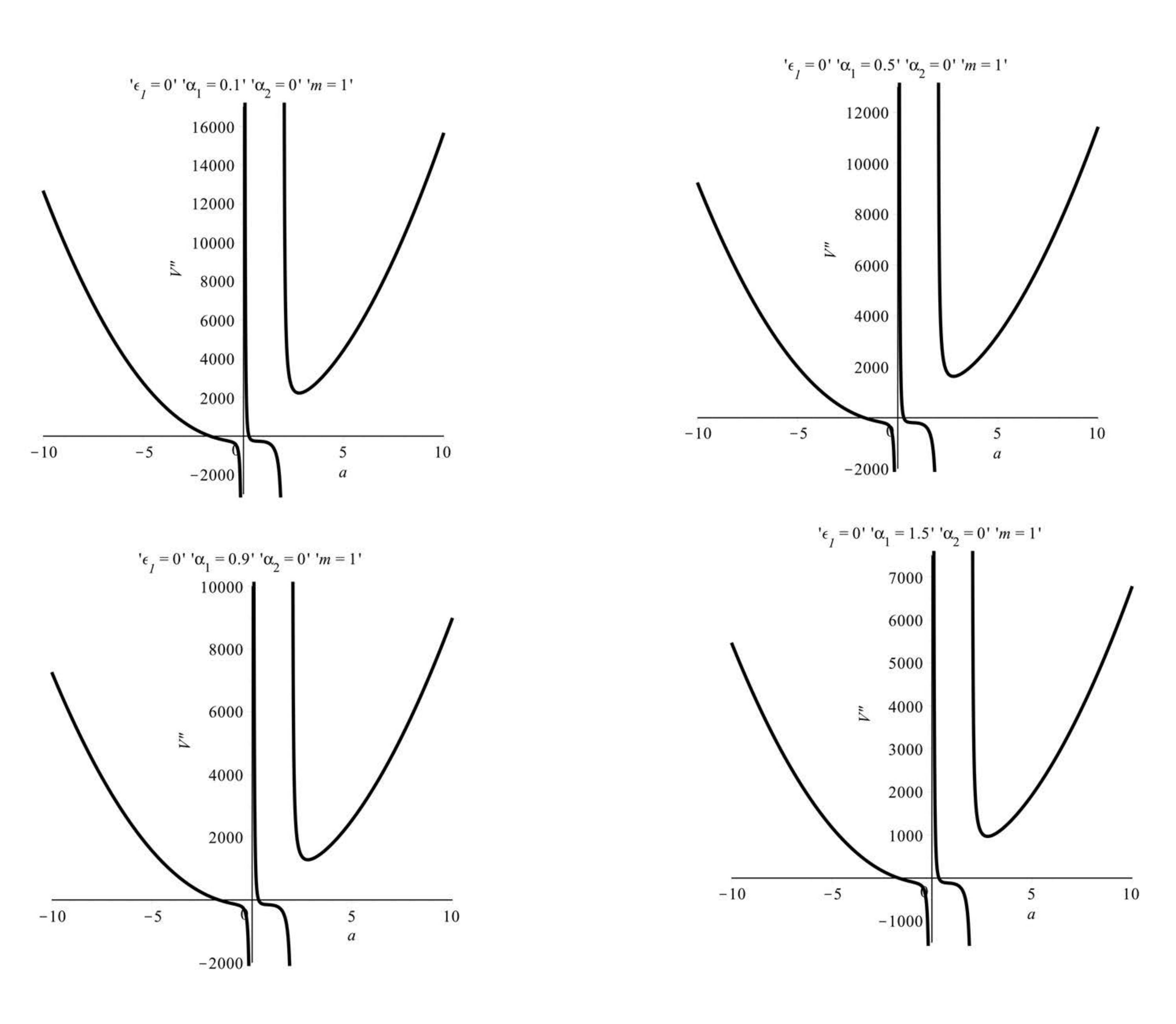} %
\caption{Here we plot the $V''$ versus $a$ to show  stability regions via the MCG }
\end{figure}

\begin{figure}[!htb]

  \includegraphics[width=4 in]{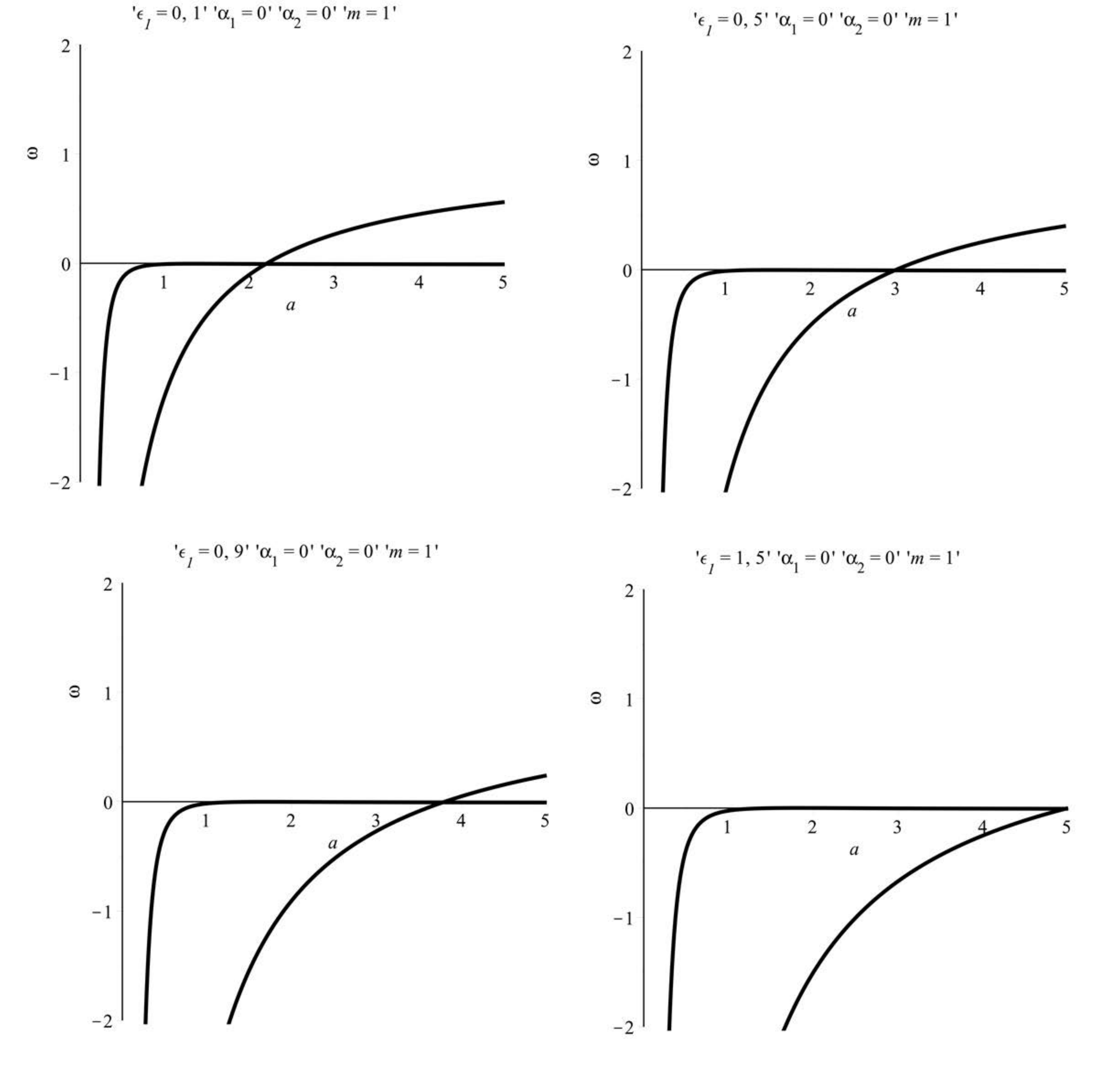} %
\caption{Here we plot the $\omega$ versus $a$ to show stability regions via the MCG }

\end{figure}

\begin{figure}[!htb]

  \includegraphics[width=4 in]{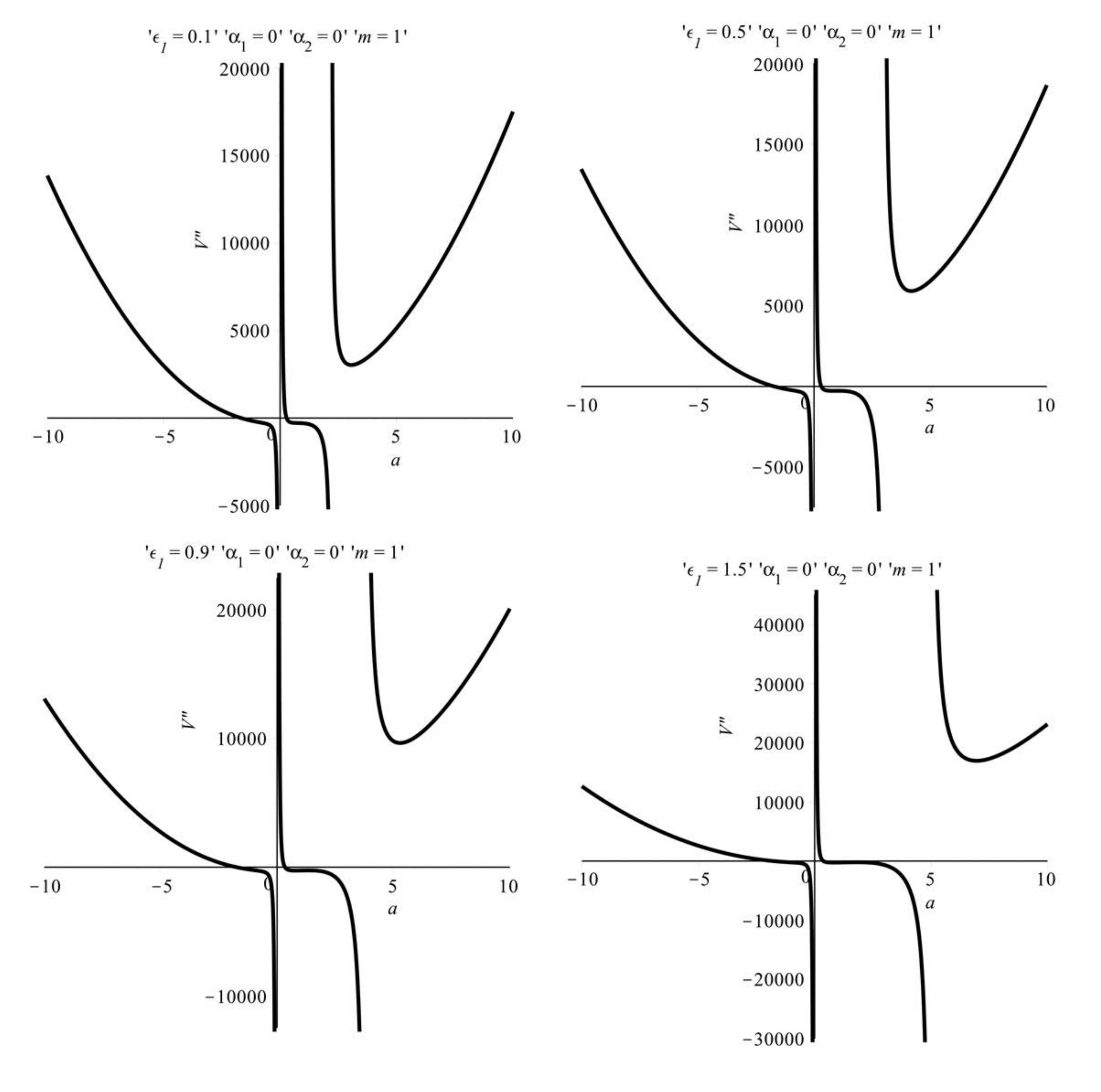} %
\caption{Here we plot the $V''$ versus $a$ to show stability regions via the MCG}

\end{figure}

\begin{figure}[!htb]

  \includegraphics[width=4 in]{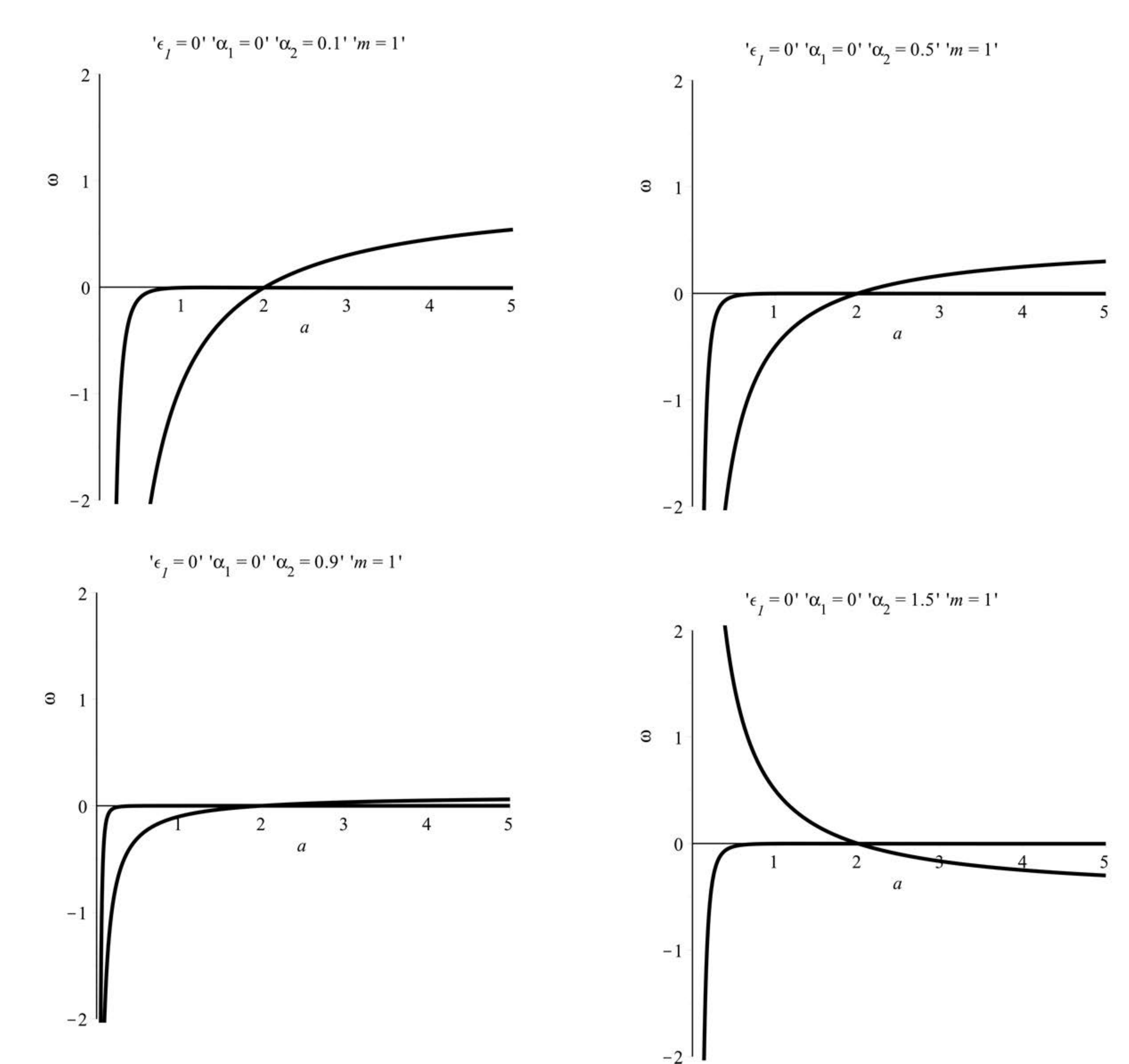} %
\caption{Here we plot the $\omega$ versus $a$ to show  stability regions via the MCG }
\end{figure}

\begin{figure}[!htb]

  \includegraphics[width=4 in]{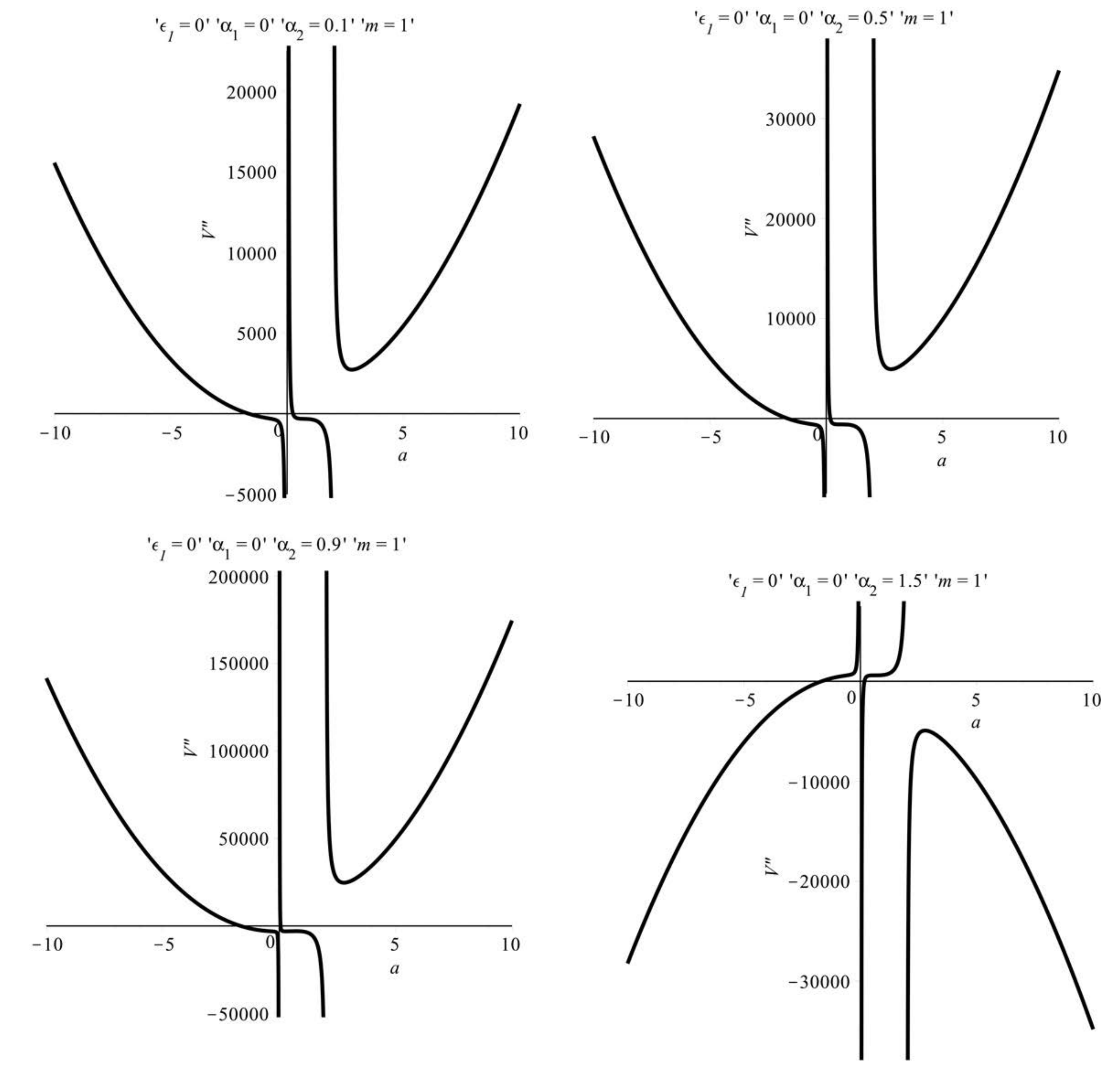} %
\caption{Here we plot the $V''$ versus $a$ to show stability regions via the MCG}
\label{fig2}
\end{figure}

\newpage
\subsection{Stability analysis of the effective TSW with a Dark Sector}

Now, we use the EOS of MCG as a unified DE and DM, which is known as a dark sector, can be cast into \cite{ref:MCG,ref:LuMCG,de}
\begin{equation}
p_{MCG}=B\rho_{MCG}-A/\rho^{\alpha}_{MCG} \label{DS}
,\end{equation} 
where $B$, $A$ and $\alpha$ are free parameters. From the Friedman-Robertson-Walkers (FRW) universe, using the energy conservation of MCG, one can rewrite the energy density of MCG as follows:
\begin{equation}
\rho_{MCG}=\rho_{MCG0}\left[B_{s}+(1-B_{s})a^{-3(1+B)(1+\alpha)}\right]^{\frac{1}{1+\alpha}},\label{eq:mcg}
\end{equation}
for $B\neq -1$, where $B_{s}=A/(1+B)\rho^{1+\alpha}_{MCG0}$. For the positive energy density, the condition $0\le B_s \le 1$ is written. The standard $\Lambda$CDM model is recovered when $\alpha=0$ and $B=0$. For the MCG with dark sector, the Friedmann equations can be written as follows \cite{de}:

\begin{eqnarray}
H^{2} & = & H_{0}^{2}\{\Omega a^{-3}+\Omega_{r}a^{-4}+\Omega_{k}a^{-2}+(1-\Omega_{b}-\Omega_{r}-\Omega_{k})\nonumber \\
 &  & \times\left[B_{s}+(1-B_{s})a^{-3(1+B)(1+\alpha)}\right]\}.
\end{eqnarray}

Note that here $H$ is the Hubble parameter and its current value is $H_{0}=70h\text{km s}^{-1}\text{Mpc}^{-1}$ \cite{ligou}, and $\Omega_{i}$ ($i=b,r,k$) are dimensionless energy parameters of baryon, radiation and effective curvature density, respectively. Furthermore, when $B=0$, the MCG becomes the generalized Chaplygin gas, which is the simplest single dark fluid model (DM + DE). Moreover, if $B=0$ and $\alpha=0$, it becomes $\Lambda$CDM model. In this paper, we consider $B \neq 0$, in which case the sound speed is given by \cite{avelino,de2}
\be
c_s^2=\frac{dp}{d\rho}=B+\alpha\frac{A}{\rho^{\alpha+1}}\,.
\ee
It is noted that imaginary part of the sound speeds are related with instabilities especially on very small scales. Here we consider a model with parameters of range $A \ge 0$, $B \ge  0$ and $0 \le \alpha \le 1$, with $c_s^2 \ge 0$. It is noted that the value of $c_s^2$ is always larger than or equal to $B$ (for large energy densities). In this limit, the MCG behaves similarly to DM. The constraint obtained using the data of CMB \cite{de}. The $\Lambda$CDM model is obtained for small values of $\alpha$ and $B$. Furthermore, today data is favor of the MCG model.

To check the stability of the effective TSW, we rewrite Eq.(\ref{DS}) as follows: \cite{de,de2,avelino}
\begin{equation}
\psi=\beta \sigma_{0} - \left(\frac{\eta}{\sigma_0^{\zeta}}\right),
\end{equation}
where $B=\beta ,\rho_{MCG}=\sigma_{0},A=\eta,\alpha=\zeta$. The resulting first derivative of function $\psi$ is
\begin{equation}
\psi^{\prime}(\sigma_{0})=\beta+\zeta \left(\frac{\eta}{\sigma_{0}^{\zeta+1}}\right).
\end{equation}

Note that $\beta$, $\eta$ and $\zeta$ are constant parameters.

To show the stability regions of the effective TSW with MCG as a unified DM and DE models, we plot the second derivative of the potential $V''$ versus $a$ in Fig. (9). Here we use EOS of the DE, then one can naturally determine the range of stable junction radius in \cite{wangm}.

\begin{figure}[!htb]

  \includegraphics[width=2.5 in]{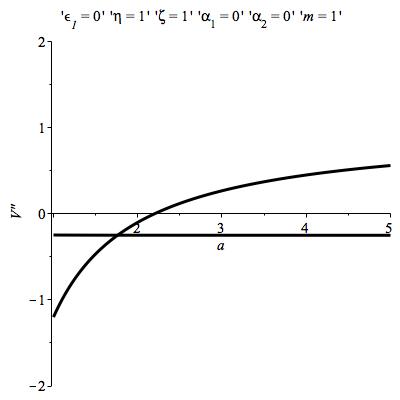} %
  \includegraphics[width=2.5 in]{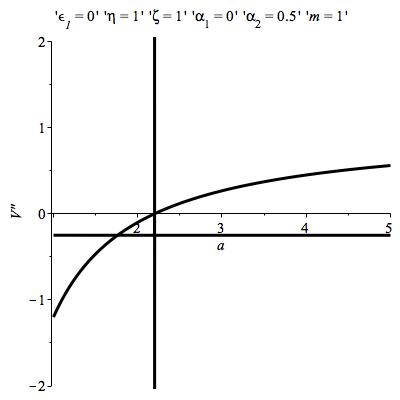} %
  \includegraphics[width=2.5 in]{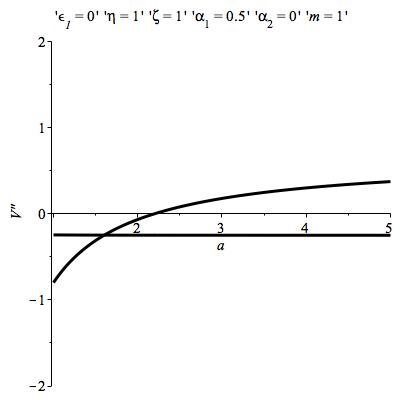} %
  \includegraphics[width=2.5 in]{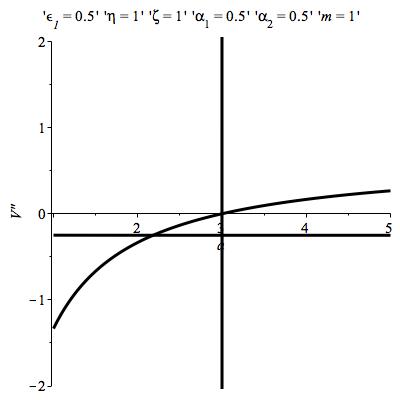} %
\caption{Here we plot the $V''$ versus $a$ to show stability regions with MCG as DE and DM with $\beta=1$}
\label{fig12}
\end{figure}

\begin{figure}[!htb]

   \includegraphics[width=2.5 in]{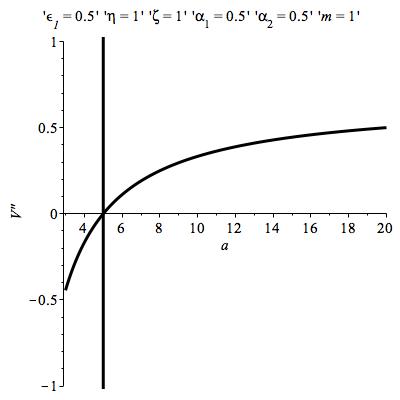} %
  \includegraphics[width=2.5 in]{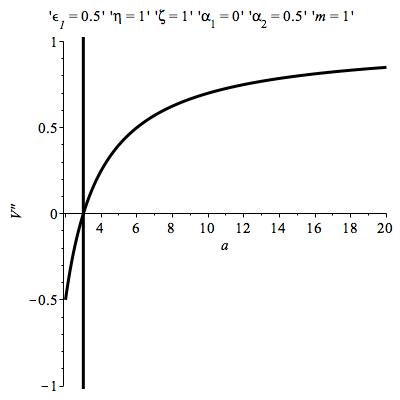}
  \includegraphics[width=2.5 in]{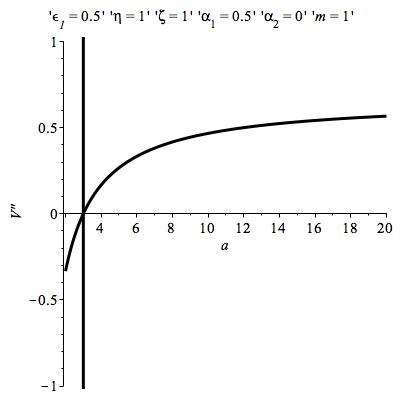}
  \includegraphics[width=2.5 in]{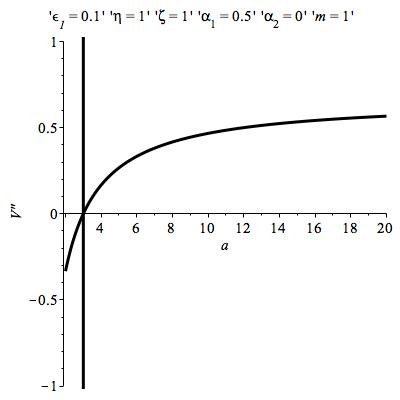}
\caption{Here we plot the $V''$ versus $a$ to show stability regions with MCG including only DE according to $1\sigma$ range of dark energy EoS $\omega$ with $\beta=1$ \cite{wangm} }
\label{fig13}
\end{figure}

\newpage

\section{Conclusion}
In this paper, we have constructed a more general TSW--namely, a generic spherically symmetric TSW solution using the Darmois\textendash Israel formalism. Next, we have considered a typical example by constructing an effective TSW with a MCG and a DS as a unified DM and DE model by choosing data from the combination of the full CMB, BAO and SN Ia data points. In contrast to the reports in the literature, we use MCG as an entire energy component (DM+DE) without any decomposition. Hence, we show that it is possible to find stable regions, and it is concluded that $\alpha_{1}$ and $\alpha_{2}$ are the most critical factors for the existence of such a stable effective TSW.

\begin{acknowledgments}
This work was supported by the Chilean FONDECYT Grant No. 3170035 (A\"{O}). A\"{O} is grateful to the CERN theory (CERN-TH) division for hospitality where part of this work was done.
\end{acknowledgments}

\end{document}